\documentclass[aps,pra,twocolumn,showpacs,superscriptaddress]{revtex4-2}
\usepackage[usenames,dvipsnames,svgnames,table]{xcolor}
\usepackage{amssymb,amsmath,amsfonts}
\usepackage[stable]{footmisc}
\usepackage{tabularray}
\usepackage{hyperref}
\usepackage{mathrsfs}
\usepackage{bbm,bm}
\usepackage{circuitikz}
\usepackage{tikz}
\usetikzlibrary{arrows,fit,positioning,automata,arrows.meta}
\usetikzlibrary{decorations,decorations.pathmorphing,decorations.pathreplacing,shapes.arrows,shapes.geometric}

\definecolor{light-gray}{gray}{0.7}

\newenvironment{definition}[1][Definition]{\begin{trivlist}
\item[\hskip \labelsep {\bfseries #1}]}{\end{trivlist}}

\newcommand{\ket}[1]{\left| #1 \right>} 
\newcommand{\bra}[1]{\left< #1 \right|} 

\begin{document}

\title{Quantum autonomous Boolean networks}

\author{Ian T. Durham}
\email[]{idurham@anselm.edu}
\affiliation{Department of Physics, Saint Anselm College, Manchester, NH 03102, USA}
\date{\today}

\begin{abstract}
Boolean networks, first developed in the late 1960s as a tool for studying complex disordered dynamical systems, consist of nodes governed by Boolean functions whose evolution is entirely deterministic in that the state of the network at a given time fully determines the state of the network at some future time. They are known for exhibiting a high degree of spontaneous order and have since become a fundamental tool for modeling a wide variety of systems. In this article I develop a model for quantum autonomous Boolean networks that exhibits many of the same properties as the classical model while also demonstrating uniquely quantum properties within a rich landscape of behavior.
\end{abstract}
\maketitle

\section{Introduction}
\label{sec:intro}

Boolean networks were first developed by Kauffman in the late 1960s as a tool for studying complex disordered dynamical systems~\cite{Kauffman:1969aa}. His original intent was to find networks that might possess enough order such that they allow for adaptation and selection as in, for example, genetic regulatory processes and similar systems~\cite{Kauffman:1990aa}. The nodes of these networks consist of Boolean functions and their evolution is entirely deterministic; the state of the network at a given time $t$ fully determines the state of the network at time $t+1$. As such, these networks are referred to as \textit{autonomous}. If the initial connections between the nodes are randomly determined, then these networks are also referred to as \textit{random} Boolean networks. Because these networks have a finite number of nodes and are entirely deterministic, they eventually cycle through a finite number of states, i.e. they exhibit state cycles of finite length where the length depends on the number of functions in the network. In some of these networks, the state of one or more of the nodes are ``frozen'' in a given state regardless of the length of the state cycle. Such elements are referred to as \textit{frozen cores}. The frozen cores create ``islands'' of isolated nodes separated by ``percolating walls'' such that perturbations to variables in one island have no effect on variables in other islands~\cite{Kauffman:1990ab}. Since Kauffman's introduction, the properties of these networks have been extensively studied~\cite{Bastolla:1998,Bilke:2001,Socolar:2003,Samuelsson:2003,Aldana:2003,Mihaljev:2006,Drossel:2009,Zhang:2009,S.-Cavalcante:2010aa,Greil:2012}. They are considered one of the fundamental modeling tools in the study of complex biological systems~\cite{Schwab:2020} and are also considered an important tool in the modeling of stochastic dynamical systems~\cite{Klemm:2005,Mozeika:2012,Jansen:2013}. They serve as the canonical example of mechanisms in integrated information theory and its quantum generalizations~\cite{Oizumi:2014aa,Zanardi:2018aa,Barbosa:2021aa,Kleiner:2021aa,Albantakis:aa}, and have also found applicability in cryptography~\cite{Wang:2020,Gao:2023}.

To date, however, little work has been done on quantum extensions or analogs of these networks. By extensions and analogs, I am not referring here to the simulation of these networks on quantum systems, but rather to more direct implementations of these networks as circuits of Boolean functions. This, of course, requires modeling the functions as unitary gates but since not all \textit{classical} Boolean functions are reversible, their implementation as unitary gates in quantum circuits often requires the use of ancilla qubits~\cite{Albantakis:aa}. A method for the construction of unitary gates for the implementation of classical Boolean functions was first proposed by Deutsch in 1985~\cite{Deutsch:1985aa}. The more general concept of \textit{quantum} Boolean functions was developed by Montanaro and Osborne in 2010~\cite{Montanaro:2010aa}. The latter was recently used as a basis for the development of a quantum analog of Kauffman's original model~\cite{Franco:2021aa}. This analog, however, left a number of fundamental issues unaddressed. In particular it did not explore the effects of quantization on a number of issues, including the lengths of state cycles and the existence of frozen cores, which are of vital importance to the emergence of order and to the robustness of these networks to random perturbations. 

This article thus develops a model framework for such networks that includes these features, allowing for a more direct comparison to Kauffman's networks and extending the realm of study for such networks to the quantum domain. The framework reveals a rich landscape of behavior unique to quantum systems including highly variable state cycle lengths that do not depend on the number of functions in the network, as well as the existence of correlations between isolated islands across frozen cores. This framework, however, is \textit{not} intended as a way to model Kauffman's networks directly using quantum systems. That is, I am not interested here in reproducing the exact sequences of states that are generated by classical autonomous Boolean networks. Rather, I am interested in the behavior of sets of Boolean functions modeled as unitary gates in the quantum domain. So, while it might be possible to recreate the properties of an autonomous Boolean network consisting of a certain set of Boolean functions using a set of quantum logic gates, instead, in the framework presented here, the network's gates would be direct quantum implementations of the those classical Boolean functions. In other words, there is a one-to-one correspondence in the present work between the number of operations in the network and the number of Boolean functions that are represented.

I begin in Section~\ref{sec:bool} with a review of Boolean functions in general before introducing their unitary implementations and a generalized bit oracle for calculating irreversible classical Boolean functions on quantum networks. In Section~\ref{sec:auto} I then give an overview of classical autonomous Boolean networks along with a discussion of circuit implementations of such networks before introducing the quantum framework. In Section~\ref{sec:qabns} I discuss the important features of these networks, which I refer to as quantum autonomous Boolean networks (qABNs), and compare them to their classical counterparts. Finally, in Section~\ref{sec:conc} I discuss some of the unanswered questions and lines of inquiry that might be undertaken to address them.

\section{Boolean functions}
\label{sec:bool}

A \textit{classical} Boolean function is a function of $k$ input variables and $m$ output variables of the form $f:\{0,1\}^k \to \{0,1\}^m$ where $\{0,1\}$ is the standard Boolean domain and is isomorphic to $\mathbbm{Z}/2\mathbbm{Z}$ where the addition of any two variables $x$ and $y$ is $x\oplus y$. In the quantum domain, each of our $k$ input variables is represented by a qubit. A \textit{quantum} Boolean function of $k$ qubits is then a unitary operator $f$ on $k$ qubits such that $f^2=\mathbbm{1}$.

It is often standard procedure to identify $\{0,1\}$ with $\{+1,-1\}$ by defining $0\equiv +1$ and $1\equiv -1$~\cite{Montanaro:2010aa}. This makes $\{+1,-1\}$, which is also isomorphic to $\mathbbm{Z}/2\mathbbm{Z}$, the multiplicative group of two elements where products are written $xy$. In this paper, unless otherwise specified, I will employ the $\{0,1\}$ convention in order to better demonstrate the consistency with Kauffman's original ideas.

Consider a single variable $x$ defined on the Boolean domain $\{0,1\}$ whose state is determined by $k$ input variables, each of which is itself defined on the Boolean domain. The number of combinations of states of $k$ inputs is just $2^k$. But for each of these $2^k$ combinations, a specific Boolean function must, by definition, specify the value of $x$. Since the values of $x$ lie on the Boolean domain, this means that there are a total of $2^{2^k}$ Boolean functions of $k$ inputs. For example there are sixteen two-input Boolean functions, including the familiar AND, OR, and NOT.

There are two natural ways of representing classical Boolean functions on quantum systems. These are known as the \textit{phase oracle}~\cite{Kashefi:2002aa}:
\begin{equation}
\ket{\mathbf{x}}\mapsto (-1)^{f(\mathbf{x})}\ket{\mathbf{x}}
\label{eq:phase}
\end{equation}
and the \textit{bit oracle} (also called the \textit{standard oracle}):
\begin{equation}
\ket{\mathbf{x}}\ket{\mathbf{y}}\mapsto U_f \ket{\mathbf{x}}\ket{\mathbf{y}}\equiv \ket{\mathbf{x}}\ket{\mathbf{y}\oplus f(\mathbf{x})}
\label{eq:bit}
\end{equation}
where $U_f^2=U_f^{\dag}U_f=\mathbbm{1}$, $\mathbf{x}\equiv x_1x_2\cdots x_k, x_j \in \{0,1\}$, and $\mathbf{y}\equiv y_1y_2\cdots y_m, y_i \in \{0,1\}$. That is, given an input state $\ket{\mathbf{x}}\ket{\mathbf{y}}$, $U_f$ maps the function's logical output to $\ket{\mathbf{y}}$.

\subsection{Unitary implementations}
\label{sec:uni}
The logical output $\ket{\mathbf{y}}$ may or may not be a set of ancillas depending on the nature of the function. If a function is naturally reversible then no ancilla is needed. But if the function is not reversible then at least one ancilla is required in order to ensure unitarity. To see this, consider the truth tables of the exclusive OR (XOR) and OR functions in Table~\ref{tab:truth}.
\begin{table}
\begin{tblr}{Q[c,m] Q[c,m] |[0.5pt] Q[c,m] Q[c,m] Q[c,m] Q[c,m] Q[c,m] |[0.5pt] Q[c,m] Q[c,m]}
\SetCell[c=4]{c} XOR & & & & & \SetCell[c=4]{c} OR & & & \\
$x$ & $y$ & $x$ & $y\oplus f(x)$ & & $x$ & $y$ & $x$ & $y\oplus f(x)$ \\ \cline[0.5pt]{1-4} \cline[0.5pt]{6-9}
0 & 0 & 0 & 0 & & 0 & 0 & 0 & 0 \\
0 & 1 & 0 & 1 & & 0 & 1 & 0 & 1 \\
1 & 0 & 1 & 1 & & 1 & 0 & 1 & 1 \\
1 & 1 & 1 & 0 & & 1 & 1 & 1 & 1
\end{tblr}
\caption{\label{tab:truth} As the truth table for the XOR function indicates, each pair of outputs can uniquely be identified with a pair of inputs which means the XOR function is reversible and thus does not require an ancilla. There is ambiguity in the OR function since both $(x,y) = (1,0)$ and $(x,y) = (1,1)$ as inputs lead to $(x,y\oplus f(x)) = (1,1)$ as an output. As such the OR function is not reversible and thus requires an ancilla in order to be represented reversibly.}
\end{table}
Every pair of outputs $(x,y\oplus f(x))$ for the XOR function is uniquely specified by a pair of inputs $(x,y)$. The same is not true for the OR function since the output pair $(x,y\oplus f(x))=(1,1)$ is obtained from two different input pairs. Thus the OR function is not reversible and an additional input and output would need to be specified in order for this function to be represented reversibly. A reversible truth table for the OR function with $y$ serving as an ancilla is shown in Table~\ref{tab:truthor} where $y$ is always assumed to start in the $0$ state.
\begin{table}
\begin{tblr}{colspec={Q[c,m] Q[c,m] Q[c,m,light-gray] |[0.5pt] Q[c,m] Q[c,m] Q[c,m,light-gray]}}
\SetCell[c=6]{c} OR & & & & & \\
$x_1$ & $x_2$ & $y$ & $x_1$ & $x_2$ & $y\oplus f(x)$ \\ \hline[0.5pt]
0 & 0 & 0 & 0 & 0 & 0 \\
0 & 1 & 0 & 0 & 1 & 1 \\
1 & 0 & 0 & 1 & 0 & 1 \\
1 & 1 & 0 & 1 & 1 & 1
\end{tblr}
\caption{\label{tab:truthor} In order to reversibly represent the OR function, $y$ must be an ancilla to which the logical output is mapped.}
\end{table}

It's clear, then, that not all classical Boolean functions of $k$ inputs can be reversibly represented with those $k$ inputs alone. This can become cumbersome for systems of multiple Boolean functions and so it might be natural to ask if there is a way to optimize the system. Suppose we have two classical Boolean functions, each of which has two logical inputs and each of which requires an ancilla in order to be represented reversibly. That is, both functions are of the form
\begin{equation}
(x_1,x_2,y)\mapsto (x_1,x_2,y\oplus f(x_1,x_2)).
\label{eq:twoin}
\end{equation}
We might be tempted to simultaneously implement these two functions as
\begin{equation}
(x_1,x_2,y)\mapsto (x_1,x_2\oplus f_1(x_1,y),y\oplus f_2(x_1,x_2))
\label{eq:multiuni}
\end{equation}
But~\ref{eq:multiuni} isn't necessarily reversible. A simple example will suffice to show this. The truth table for~\ref{eq:multiuni} with $f_1(x_1,y)$ taken to be the AND function and $f_2(x_1,x_2)$ taken to be the OR function is shown in Table~\ref{tab:truthmulti}.
\begin{table}
\begin{tblr}{Q[c,m] Q[c,m] Q[c,m] |[0.5pt] Q[c,m] Q[c,m] Q[c,m]}
\SetCell[c=4]{c} & & & & AND & OR \\
$x_1$ & $x_2$ & $y$ & $x_1$ & $x_2\oplus f_1(x_1,y)$ & $y\oplus f_2(x_1,x_2)$ \\ \hline[0.5pt]
0 & 0 & 0 & 0 & 0 & 0 \\
1 & 0 & 0 & 1 & 0 & 1 \\
0 & 1 & 0 & 0 & 0 & 1 \\
0 & 0 & 1 & 0 & 0 & 0 \\
1 & 0 & 1 & 1 & 1 & 1 \\
1 & 1 & 0 & 1 & 0 & 1 \\
0 & 1 & 1 & 0 & 0 & 1 \\
1 & 1 & 1 & 1 & 1 & 1
\end{tblr}
\caption{\label{tab:truthmulti} If we attempt to simultaneously implement both the AND and the OR function on the same set of three inputs, we find that the process is not reversible since every unique output string could have arisen from one of two possible input strings.}
\end{table} 
As the truth table shows, none of the output triples is uniquely determined by a single input triple. As such one ancilla is required for each function. For example, if $f_1(x_1,x_2)$ is the AND function and $f_2(x_1,x_2)$ is the OR function,
\begin{equation}
(x_1,x_2,y_1,y_2)\mapsto (x_1,x_2,y_1\oplus f_1(x_1,x_2),y_2\oplus f_2(x_1,x_2))
\label{eq:multianc}
\end{equation}
\textit{is} reversible.

A similar process can be used to show that for any single irreversible classical Boolean function of $k$ inputs and $m$ outputs (where I am assuming that $m\le k$ -- the problem is more complicated when $m>k$) one ancilla is required for each output in order to implement it reversibly. The total number of elements in any such string must then be $k+m$. For $n$ such functions to be implemented with the same set of logical inputs, the total number of elements in the tuple must then be $n(k+m)$. Thus for any system of $n$ irreversible Boolean functions to be implemented reversibly, a total of $n\times m$ ancillas are required. The total length of the input and output strings must then be $n(k+m)$. But there is an additional problem that must be considered.

The point of reversible representations is to allow for unitary implementations by quantum systems. But consider a simple system of just two logical qubits and two ancillas to which we apply two (unitary) functions. We might naively expect that the bit oracle allows us to write
\begin{equation*}
\ket{x_1,x_2}\ket{y_1,y_2}\mapsto U_{f_1}U_{f_2} \ket{x_1,x_2}\ket{y_1,y_2}.
\end{equation*}
But there's nothing privileging the order of the operators. We could just as easily have written
\begin{equation*}
\ket{x_1,x_2}\ket{y_1,y_2}\mapsto U_{f_2}U_{f_1} \ket{x_1,x_2}\ket{y_1,y_2}
\end{equation*}
But for these to produce the same result, the unitary representations of our functions would have to be commutative, i.e. $U_{f_1}U_{f_2}$ would have to be equal to $U_{f_2}U_{f_1}$. Of course not all unitary operators are commutative. The Pauli operators, for example, which are valid \textit{quantum} single-input Boolean functions are not commutative. However, I am primarily concerned here with those unitary operators that represent \textit{classical} Boolean functions and it remains an open question as to whether all such operators are commutative. If not, the number of qubits required to implement some of these systems could be quite large. In this article I will keep things simple and focus on Boolean functions for which $k=2$ and $m=1$ and will implement each with its own, unique set of qubits.

\subsection{A generalized bit oracle}
\label{sec:genbit}

We can generalize the notion of a bit oracle to include both pure and mixed states by defining
\begin{equation}
\rho(\mathbf{x},\mathbf{y}) \equiv \ket{\mathbf{x}}\ket{\mathbf{y}}\bra{\mathbf{y}}\bra{\mathbf{x}}
\label{eq:density}
\end{equation}
The action of a Boolean function on this state is then a unitary transformation
\begin{equation}
\rho(\mathbf{x},\mathbf{y}\oplus f(\mathbf{x})) = U_f \rho(\mathbf{x},\mathbf{y})U_f^{\dag}.
\label{eq:unitaryone}
\end{equation}
This is a generalization of the bit oracle to a broader class of states. Now consider a system of $n$ Boolean functions and define
\begin{align}
\rho(\mathbf{X},\mathbf{Y}) & \equiv \rho(\mathbf{x}_1,\mathbf{y}_1)\otimes\cdots\otimes\rho(\mathbf{x}_n,\mathbf{y}_n) \nonumber \\
U_F & \equiv U_{f_1}\otimes\cdots\otimes U_{f_n} \nonumber \\
F(\mathbf{X}) & \equiv f_1(\mathbf{x}_1)\otimes\cdots\otimes f_n(\mathbf{x}_n).
\label{eq:tensor}
\end{align}
The evolution of the full system is then
\begin{equation}
\rho(\mathbf{X},\mathbf{Y}\oplus F(\mathbf{X})) = U_F \left[\rho(\mathbf{X},\mathbf{Y})\right] U_F^{\dag}.
\label{eq:evolution}
\end{equation}
This generalizes the bit oracle to a network of $n$ classical Boolean functions and serves as the governing equation for the evolution of the network.

\section{Autonomous Boolean networks}
\label{sec:auto}
An autonomous Boolean network (ABN) is a network of $n$ Boolean variables (i.e. their values are on the Boolean domain) whose state at some time $t$ fully determines the state of the network at time $t+1$ via a set of Boolean functions acting on the set of variables. As such these networks are fully deterministic in their evolution and there are no additional external variables introduced at any point in the evolution of the network. If each variable is updated simultaneously, the network is said to be \textit{synchronous}~\cite{Kauffman:1990ab}. Only synchronous networks are considered in this article.

The networks developed by Kauffman were described as both autonomous and \textit{random}. But this latter description is misleading. It refers to how the connections between the nodes in the network are \textit{initially} set. But thereafter the connections remain fixed. Since the point of these networks is to characterize the study the general behavior of all such networks and their connections by studying some subset of them, there's really nothing random about it. It's simply a way to choose which network to study at a given moment. As such I will refrain from referring to these networks as random.

\subsection{Classical networks}
\label{sec:classical}
Consider a simple network of three variables, $x_1,x_2,x_3 \in\{0,1\}$, each of which receives inputs from the other two. Assume that the dynamical evolution of the first is governed by the AND function and the dynamical evolution of the other two are governed by the OR function. The truth table for this simple network is shown in Table~\ref{tab:truthtri} where, for simplicity, I have set $y = x_3$.
\begin{table}
\begin{tblr}{Q[c,m] Q[c,m] Q[c,m] |[0.5pt] Q[c,m] Q[c,m] Q[c,m]}
\SetCell[c=3]{c} & & & AND & OR & OR \\
\SetCell[c=3]{c} & & & $x_1\oplus$ & $x_2\oplus$ & $x_3\oplus$ \\
$x_1$ & $x_2$ & $x_3$ & $f_1(x_2,x_3) $ & $f_2(x_1,x_3)$ & $f_3(x_1,x_2)$ \\ \hline[0.5pt]
0 & 0 & 0 & 0 & 0 & 0 \\
1 & 0 & 0 & 0 & 1 & 1 \\
0 & 1 & 0 & 0 & 0 & 1 \\
0 & 0 & 1 & 0 & 1 & 0 \\
1 & 0 & 1 & 0 & 1 & 1 \\
1 & 1 & 0 & 0 & 1 & 1 \\
0 & 1 & 1 & 1 & 1 & 1 \\
1 & 1 & 1 & 1 & 1 & 1
\end{tblr}
\caption{\label{tab:truthtri} The truth table for a classical Boolean network of three variables whose values are governed by the AND, OR, and OR functions respectively is shown here.}
\end{table} 
This, of course, is not reversible but serves as a useful example for demonstrating the basic properties of these networks. 

The first thing to notice is that, since there are a finite number of states and the evolution is entirely deterministic, the system will eventually pass through a given state more than once. In fact it will continue to cycle through the same set of states, referred to as a \textit{state cycle}, ad infinitum. The state cycles themselves are referred to as the \textit{dynamical attractors} of the network and the set of all states leading into or lying on a given cycle are said to constitute the \textit{basin of attraction}. The \textit{length} of a given state cycle is the number of states on that cycle and can range from unity for a steady state up to $2^n$ depending, in part, on the number of inputs $k$ to each function. The basins of attraction partition the $2^n$ state space of the network. 

\begin{figure}
\begin{center}
\begin{tikzpicture}
\node at (0,3) {000};
\draw[-{Latex[length=0.75mm,width=0.75mm]}] (0.25,3) arc (-90:180:1.5mm and 1.5mm);

\node at (-0.75,2) {001};
\draw[{Latex[length=0.75mm,width=0.75mm]}-{Latex[length=0.75mm,width=0.75mm]}] (-0.4,2) -- (0.4,2);
\node at (0.75,2) {010};

\node at (0,0) {011};
\node at (-1.25,0) {110};
\draw[-{Latex[length=0.75mm,width=0.75mm]}] (-0.9,0) -- (-0.35,0);
\node at (1.25,0) {111};
\draw[-{Latex[length=0.75mm,width=0.75mm]}] (0.35,0) -- (0.9,0);
\node at (0,1) {100};
\draw[-{Latex[length=0.75mm,width=0.75mm]}] (0,0.75) -- (0,0.25);
\node at (0,-1) {101};
\draw[-{Latex[length=0.75mm,width=0.75mm]}] (0,-0.75) -- (0,-0.25);
\draw[-{Latex[length=0.75mm,width=0.75mm]}] (1.5,0) arc (-90:180:1.5mm and 1.5mm);

\node at (3.5,3) {state cycle 1};

\node at (3.5,2) {state cycle 2};

\node at (3.5,0) {state cycle 3};

\end{tikzpicture}
\end{center}
\caption{\label{fig:statecycle} The state cycles and basins of attraction are shown for an autonomous Boolean network corresponding to the truth table in Table~\ref{tab:truthtri}. Note that the length of the third state cycle is just unity since, once it settles into the $(1,1,1)$ state, it remains there.}
\end{figure}
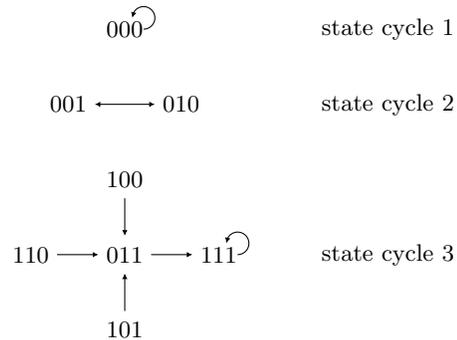

Figure~\ref{fig:statecycle} shows all the state cycles and basins of attraction for the network whose truth table is shown in Table~\ref{tab:truthtri}.
In general, the length of the state cycle and the number of attractors are both functions of the number of inputs per function $k$ and the number of functions $n$. For example, for functions with $k=2$ inputs such as the AND and OR functions, the expected state cycle length and the mean attractor length is on the order of $\sqrt{n}$, though some systems show a power law relation~\cite{Bastolla:1998,Mihaljev:2006}. This is a rather remarkable result. A system of 10,000 binary variables with $2^{10,000}$ possible states will settle and cycle through a mere 100 of these states. Additionally each state cycle (in the $k=2$ case) is stable to almost all minimal perturbations including the deletion of elements~\cite{Kauffman:1990ab}. For the network in Figure~\ref{fig:statecycle}, the second state cycle is actually the longest since, in the third basin of attraction, once the state settles into the $(1,1,1)$ state, it remains there.

Now consider a more complicated network of seven variables, each of which dynamically evolves according to some Boolean function. The variables are connected to one another such that each variable's state at time $t+1$ is determined by the states of two other \textit{not necessarily neighboring} variables at time $t$. Suppose the evolution of the first seven elements of the network over ten time steps proceeds as shown in Table~\ref{tab:walls}.
\begin{table}
\begin{tblr}{Q[c,m] |[0.5pt] Q[c,m] Q[c,m] Q[c,m] Q[c,m,light-gray] Q[c,m] Q[c,m] Q[c,m]}
$t$ & $x_1$ & $x_2$ & $x_3$ & $x_4$ & $x_5$ & $x_6$ & $x_7$ \\ \hline[0.5pt]
1 & 0 & 1 & 0 & 1 & 0 & 0 & 1 \\
2 & 1 & 0 & 1 & 1 & 1 & 1 & 0 \\
3 & 0 & 1 & 1 & 1 & 1 & 0 & 1 \\
4 & 1 & 1 & 1 & 1 & 0 & 1 & 0 \\
5 & 0 & 1 & 0 & 1 & 1 & 1 & 1 \\
6 & 1 & 0 & 0 & 1 & 1 & 0 & 0 \\
7 & 0 & 0 & 1 & 1 & 0 & 1 & 1 \\
8 & 1 & 0 & 1 & 1 & 0 & 1 & 0 \\
9 & 1 & 1 & 0 & 1 & 1 & 1 & 1 \\
10 & 0 & 1 & 0 & 1 & 1 & 0 & 1
\end{tblr}
\caption{\label{tab:walls} The evolution of the first seven variables in a hypothetical seven-variable network shows a frozen core corresponding to the fourth element which gives two functionally isolated islands consisting of ($x_1,x_2,x_3$) and ($x_5,x_6,x_7$) respectively.}
\end{table} 
Notice that the fourth variable never changes. We refer to this variable as a \textit{frozen core}; it separates all the variables to its left from those to its right. The two sides of the frozen core form \textit{functionally isolated} islands separated by a \textit{percolating wall}. Note that this does \textit{not} mean that there aren't connections between the islands. For example, it could be that the second variable's state is determined by variables one and six, i.e. $x_3\equiv x_3\oplus f(x_1,x_6)$. Nevertheless, the islands are said to be functionally isolated because \textit{perturbations to variables in one island have no effect on variables in other islands even though the islands may be connected}. As I will show, this is \textit{not} true in the quantum case. I refer the interested reader back to Kauffman for additional details on these structures~\cite{Kauffman:1990aa,Kauffman:1990ab}.

\subsection{Circuit implementations}
\label{sec:circuits}
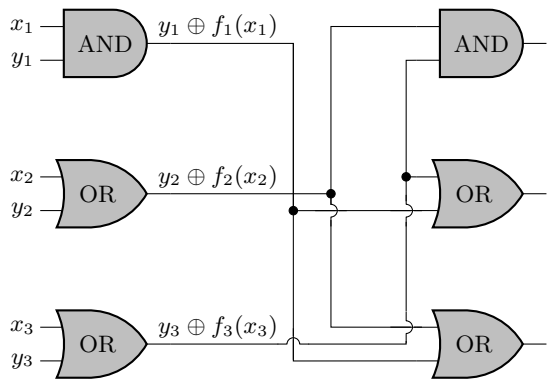
\begin{figure}
\begin{center}
\begin{tikzpicture}
\ctikzset{
    logic ports=ieee,
    logic ports/scale=0.8,
    logic ports/fill=lightgray
}
\node[and port] (ANDa) at (-1,0){};
\node at (-1,0) {AND};
\node[or port] (ORa) at (-1,-2){};
\node at (-1.1,-2) {OR};
\node[or port] (ORb) at (-1,-4){};
\node at (-1.1,-4) {OR};

\node[and port] (ANDb) at (4,0){};
\node at (4,0) {AND};
\node[or port] (ORc) at (4,-2){};
\node at (3.9,-2) {OR};
\node[or port] (ORd) at (4,-4){};
\node at (3.9,-4) {OR};

\node at (-2.1,0.225) {$x_1$};
\node at (-2.1,-0.225) {$y_1$};

\node at (-2.1,-1.775) {$x_2$};
\node at (-2.1,-2.225) {$y_2$};

\node at (-2.1,-3.775) {$x_3$};
\node at (-2.1,-4.225) {$y_3$};

\node at (0.5,0.225) {$y_1\oplus f_1(x_1)$};

\node at (0.5,-1.775) {$y_2\oplus f_2(x_2)$};

\node at (0.5,-3.775) {$y_3\oplus f_3(x_3)$};

\draw (ANDa.out) -- (1.5,0) -- (1.5,-4.225) -- (ORd.in 2);
\draw (1.5,0) to[short,-*] (1.5,-2.225) -- (ORc.in 2); 

\draw (ORa.out) to[short,-*] (2,-2);
\draw (2,-2) -- (2,0.225) -- (ANDb.in 1);
\draw (2,-2) -- (2,-2.15);
\node at (2,-2.225) [jump crossing,rotate=-90,scale=1.5]{};
\draw (2,-2.35) -- (2,-3.775) -- (ORd.in 1);

\draw (ORb.out) -- (1.3,-4);
\node at (1.5,-4) [jump crossing,scale=1.5]{};
\draw (1.7,-4) -- (3,-4) -- (3,-3.96);
\node at (3,-3.775) [jump crossing,rotate=-90,scale=1.5]{};
\draw (3,-3.56) -- (3,-2.375);
\node at (3,-2.225) [jump crossing,rotate=-90,scale=1.5]{};
\draw (3,-2.1) to[short,-*] (3,-1.775) -- (ORc.in 1);
\draw (3,-1.775) -- (3,-0.225) -- (ANDb.in 2);

\end{tikzpicture}
\end{center}
\caption{\label{fig:classcirc} The classical circuit diagram for a network consisting of an AND gate and two OR gates highlights two important facts: (i) the non-logical outputs ($x_1,x_2,x_3$) are discarded after the first step and (ii) each logical output ($y_1\oplus f_1(x_1),y_2\oplus f_2(x_2),y_3\oplus f_3(x_3)$) is copied so it can be used as the input to more than one function at the next step. Note that this is only one of thirty-six possible wiring diagrams for a network consisting of these three functions.}
\end{figure}

In circuit implementations of these networks there are two things worthy of note. These are best seen by considering the idealized circuit diagram for the network described in Table~\ref{tab:truthtri} as shown in Figure~\ref{fig:classcirc}.
The first thing to notice is that the bits that do not represent the logical output are discarded in the sense that they are not used to compute the network's next state. Only the logical output bits are used to compute the next step. But that, then, necessitates the second noteworthy attribute of these networks. In order to ensure the correct number of inputs on subsequent steps, the logical outputs must be copied. 

This presents a problem for any attempt to directly implement such a network on a quantum system. Since the system is quantum its evolution should be unitary. But it is well-known that no single universal unitary gate can copy (clone) an arbitrary quantum state~\cite{Park:1970aa,Wootters:1982aa,Dieks:1982aa}. As such, we can't directly implement such a network on a quantum system without additional inputs.

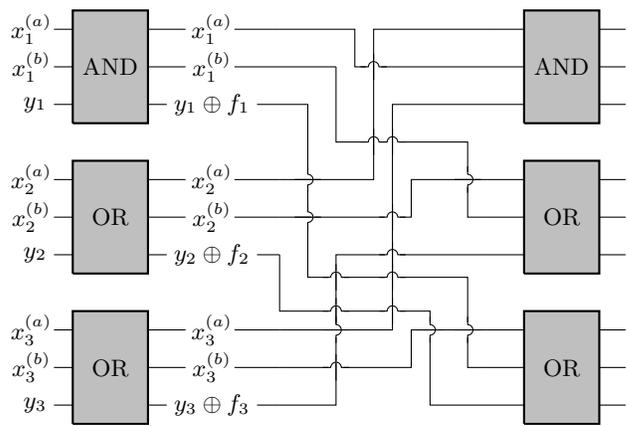
\begin{figure}
\begin{center}
\begin{tikzpicture}
\fill[lightgray] (0,0) -- (1,0) -- (1,1.5) -- (0,1.5) -- (0,0);
\draw[thick] (0,0) -- (1,0) -- (1,1.5) -- (0,1.5) -- (0,0);
\draw (-0.25,0.25) -- (0,0.25);
\draw (-0.25,0.75) -- (0,0.75);
\draw (-0.25,1.25) -- (0,1.25);
\node at (0.5,0.75) {OR};

\node at (-0.55,1.25) {$x_3^{(a)}$};
\node at (-0.55,0.75) {$x_3^{(b)}$};
\node at (-0.5,0.25) {$y_3$};

\draw (1,0.25) -- (1.25,0.25);
\node at (1.85,0.25) {$y_3 \oplus f_3$};
\draw (2.45,0.25) -- (2.75,0.25);

\draw (1,0.75) -- (1.5,0.75);
\node at (1.85,0.75) {$x_3^{(b)}$};
\draw (2.15,0.75) -- (2.75,0.75);

\draw (1,1.25) -- (1.5,1.25);
\node at (1.85,1.25) {$x_3^{(a)}$};
\draw (2.15,1.25) -- (2.75,1.25);

\fill[lightgray] (0,2) -- (1,2) -- (1,3.5) -- (0,3.5) -- (0,2);
\draw[thick] (0,2) -- (1,2) -- (1,3.5) -- (0,3.5) -- (0,2);
\draw (-0.25,2.25) -- (0,2.25);
\draw (-0.25,2.75) -- (0,2.75);
\draw (-0.25,3.25) -- (0,3.25);
\node at (0.5,2.75) {OR};

\node at (-0.55,3.25) {$x_2^{(a)}$};
\node at (-0.55,2.75) {$x_2^{(b)}$};
\node at (-0.5,2.25) {$y_2$};

\draw (1,2.25) -- (1.25,2.25);
\node at (1.85,2.25) {$y_2 \oplus f_2$};
\draw (2.45,2.25) -- (2.75,2.25);

\draw (1,2.75) -- (1.5,2.75);
\node at (1.85,2.75) {$x_2^{(b)}$};
\draw (2.15,2.75) -- (2.75,2.75);

\draw (1,3.25) -- (1.5,3.25);
\node at (1.85,3.25) {$x_2^{(a)}$};
\draw (2.15,3.25) -- (2.75,3.25);

\fill[lightgray] (0,4) -- (1,4) -- (1,5.5) -- (0,5.5) -- (0,4);
\draw[thick] (0,4) -- (1,4) -- (1,5.5) -- (0,5.5) -- (0,4);
\draw (-0.25,4.25) -- (0,4.25);
\draw (-0.25,4.75) -- (0,4.75);
\draw (-0.25,5.25) -- (0,5.25);
\node at (0.5,4.75) {AND};

\node at (-0.55,5.25) {$x_1^{(a)}$};
\node at (-0.55,4.75) {$x_1^{(b)}$};
\node at (-0.5,4.25) {$y_1$};

\draw (1,4.25) -- (1.25,4.25);
\node at (1.85,4.25) {$y_1 \oplus f_1$};
\draw (2.45,4.25) -- (2.75,4.25);

\draw (1,4.75) -- (1.5,4.75);
\node at (1.85,4.75) {$x_1^{(b)}$};
\draw (2.15,4.75) -- (2.75,4.75);

\draw (1,5.25) -- (1.5,5.25);
\node at (1.85,5.25) {$x_1^{(a)}$};
\draw (2.15,5.25) -- (2.75,5.25);

\fill[lightgray] (6,0) -- (7,0) -- (7,1.5) -- (6,1.5) -- (6,0);
\draw[thick] (6,0) -- (7,0) -- (7,1.5) -- (6,1.5) -- (6,0);
\draw (5.5,0.25) -- (6,0.25);
\draw (5.5,0.75) -- (6,0.75);
\draw (5.5,1.25) -- (6,1.25);
\node at (6.5,0.75) {OR};
\draw (7,0.25) -- (7.35,0.25);
\draw (7,0.75) -- (7.35,0.75);
\draw (7,1.25) -- (7.35,1.25);

\fill[lightgray] (6,2) -- (7,2) -- (7,3.5) -- (6,3.5) -- (6,2);
\draw[thick] (6,2) -- (7,2) -- (7,3.5) -- (6,3.5) -- (6,2);
\draw (5.5,2.25) -- (6,2.25);
\draw (5.5,2.75) -- (6,2.75);
\draw (5.5,3.25) -- (6,3.25);
\node at (6.5,2.75) {OR};
\draw (7,2.25) -- (7.35,2.25);
\draw (7,2.75) -- (7.35,2.75);
\draw (7,3.25) -- (7.35,3.25);

\fill[lightgray] (6,4) -- (7,4) -- (7,5.5) -- (6,5.5) -- (6,4);
\draw[thick] (6,4) -- (7,4) -- (7,5.5) -- (6,5.5) -- (6,4);
\draw (5.5,4.25) -- (6,4.25);
\draw (5.5,4.75) -- (6,4.75);
\draw (5.5,5.25) -- (6,5.25);
\node at (6.5,4.75) {AND};
\draw (7,4.25) -- (7.35,4.25);
\draw (7,4.75) -- (7.35,4.75);
\draw (7,5.25) -- (7.35,5.25);

\draw (2.75,3.25) -- (4,3.25) -- (4,5.25) -- (5.5,5.25);
\node at (4,4.75) [jump crossing,scale=1.25]{};

\draw (4.1,4.75) -- (5.5,4.75);
\draw (2.75,5.25) -- (3.75,5.25) -- (3.75,4.75) -- (3.9,4.75);

\draw (2.75,1.25) -- (4.25,1.25) -- (4.25,4.25) -- (5.5,4.25);
\draw (2.75,2.75) -- (4.15,2.75);
\node at (4.25,2.75) [jump crossing,scale=1.25]{};
\draw (4.35,2.75) -- (4.5,2.75) -- (4.5,3.25) -- (5.5,3.25);

\draw (2.75,4.75) -- (3.5,4.75);
\node at (4.25,3.75) [jump crossing,scale=1.25]{};
\node at (4,3.75) [jump crossing,scale=1.25]{};
\draw (3.5,4.75) -- (3.5,3.75) -- (3.9,3.75);
\draw (4.35,3.75) -- (5.25,3.75) -- (5.25,3.35);
\node at (5.25,3.25) [jump crossing,rotate=-90,scale=1.25]{};
\draw (5.25,3.15) -- (5.25,2.75) -- (5.5,2.75);

\draw (2.75,0.25) -- (3.5,0.25) -- (3.5,1.15);
\node at (3.5,1.25) [jump crossing,rotate=-90,scale=1.25]{};
\draw (3.5,1.35) -- (3.5,2.25) -- (4.15,2.25);
\node at (4.25,2.25) [jump crossing,scale=1.25]{};
\draw (4.35,2.25) -- (5.5,2.25);

\draw (2.75,0.75) -- (3.4,0.75);
\node at (3.5,0.75) [jump crossing,scale=1.25]{};
\draw (3.6,0.75) -- (4.5,0.75) -- (4.5,1.25) -- (5.5,1.25);

\draw (2.75,2.25) -- (2.75,1.5) -- (3.4,1.5);
\node at (4.25,1.5) [jump crossing,scale=1.25]{};
\draw (3.6,1.5) -- (4.15,1.5);
\node at (3.5,1.5) [jump crossing,scale=1.25]{};
\draw (4.35,1.5) -- (4.75,1.5) -- (4.75,1.35);
\node at (4.75,1.25) [jump crossing,rotate=-90,scale=1.25]{};
\draw (4.75,1.15) -- (4.75,0.25) -- (5.5,0.25);

\draw (2.75,4.25) -- (3.125,4.25) -- (3.125,3.35);
\node at (3.125,3.25) [jump crossing,rotate=-90,scale=1.25]{};
\draw (3.125,3.15) -- (3.125,2.85);
\node at (3.125,2.75) [jump crossing,rotate=-90,scale=1.25]{};
\draw (3.125,2.65) -- (3.125,1.95) -- (3.4,1.95);
\node at (4.25,1.95) [jump crossing,scale=1.25]{};
\draw (3.6,1.95) -- (4.15,1.95);
\node at (3.5,1.95) [jump crossing,scale=1.25]{};
\draw (4.35,1.95) -- (5.25,1.95) -- (5.25,1.35);
\node at (5.25,1.25) [jump crossing,rotate=-90,scale=1.25]{};
\draw (5.25,1.15) -- (5.25,0.75) -- (5.5,0.75);

\end{tikzpicture}
\end{center}
\caption{\label{fig:quantcirc} One potential wiring diagram for a quantum implementation of a Boolean network consisting of an AND and two OR functions is shown. Since there are nine outputs and nine inputs, there are eighty-one possible wiring diagrams for this set of functions. In this particular case, for example, the output $y_1 \oplus f_1(x_1^{(a)},x_1^{(b)})$ at one step is mapped to the input $x_3^{(b)}$ for the next step. Likewise, the output $y_2 \oplus f_2(x_2^{(a)},x_2^{(b)})$ at one step is mapped to the input $y_3$ which is an ancilla for the next step.}
\end{figure}

Consider a network consisting of the same functions as those just described in Figure~\ref{fig:classcirc} but with each function implemented unitarily. Neither the AND nor the OR function is naturally reversible and so each requires an ancilla qubit. The network is thus composed of nine qubits. Since we can't arbitrarily clone any of the qubits, there must be a one-to-one correspondence between the outputs at one step and the inputs at the next step. But as long as that constraint is satisfied, we can connect (wire) them in any manner we like. An arbitrary wiring diagram for this network is shown in Figure~\ref{fig:quantcirc}.
The arbitrary manner in which a given connection (wiring) was chosen is the reason Kauffman referred to them as random. The randomness was simply how a given connection was selected for study any a given time.

Notice in the circuit of Figure~\ref{fig:quantcirc} that the output $y_2 \oplus f_2(x_2^{(a)},x_2^{(b)})$ from the first step acts as the input to $y_3$ at the next step. That is, if we indicate the input to the second step with a prime, then $y_3^{\prime}\equiv y_2 \oplus f_2(x_2^{(a)},x_2^{(b)})$. But $y_3$ is an ancilla that encodes the logical output of $f_3(x_3^{\prime(a)},x_3^{\prime(b)})$. This means that the output of the second step is $y_3^{\prime} \oplus f_3(x_3^{\prime(a)},x_3^{\prime(b)}) \equiv y_2 \oplus f_2(x_2^{(a)},x_2^{(b)})\oplus f_3(x_3^{\prime(a)},x_3^{\prime(b)})$. It is usually convention to set ancillas to 0 but since there is no action external to the network here (since it is autonomous) it is entirely possible that $y_3^{\prime}\equiv y_2 \oplus f_2(x_2^{(a)},x_2^{(b)})$ will be 1. Since the action of the function is modulo 2 the logical output of the second step could produce a 0 when a 1 is expected, and vice-versa. 

As a simple example of this, consider a single OR function implemented unitarily and assume the initial input state is $(x_1=1,x_2=1,y=0)$ (corresponding to the bottom line of Table~\ref{tab:truthor}). Also assume that this network has the simplest wiring diagram such that, for example, $x_1^{\prime(a)}\equiv x_1^{(a)}$, $x_1^{\prime(b)}\equiv x_1^{(b)}$, and $y_1^{\prime}\equiv y_1$. Since the function itself is mapped to the ancilla, i.e. $y_1 = f( x_1^{(a)}, x_1^{(b)})$, the output state is $(1,1,1)$. But if we implement this function again with $(1,1,1)$ now serving as the input, the function's action on the ancilla (which is now in the state 1) is modulo 2 and thus the output becomes $(1,1,0)$. That is, for a network consisting of just the OR function and three qubits, if at any time the state of the system is either $(1,1,0)$ or $(1,1,1)$, then the system will simply cycle between these two states, i.e. $\cdots(1,1,0)\mapsto(1,1,1)\mapsto(1,1,0)\mapsto(1,1,1)\mapsto(1,1,0)\cdots$. We could choose to perform error correction at each step (that is, immediately following the output) and reset the ancilla to 0 if need be, but that would either require external intervention, in which case the network wouldn't truly be autonomous, or would consist of more than just the functions that defined the network, i.e. it would require additional operations. Since I am primarily interested here in the autonomous behavior of sets of Boolean functions, I will not execute any error correction on the ancilla qubits.

\section{Quantum networks}
\label{sec:quantum}
I am now in a position to define a quantum autonomous Boolean network (qABN) in direct analogy to classical autonomous Boolean networks.
\begin{definition}[Definition 1]
A quantum autonomous Boolean network is a network of $n$ Boolean functions of $k$ logical qubits whose state at time $t$ determines, through unitary evolution via a generalized bit oracle, its state at time $t+1$ and for which each output qubit at time $t$ maps to some input qubit (not necessarily itself) at time $t+1$, i.e. there is a one-to-one correspondence between the number of outputs at one step and the number of inputs at the next step.
\end{definition}
In order to ensure unitarity, each function can act on up to $k+m$ \textit{total} qubits where $m$ is the maximum number of ancillas required. These networks thus consist of up to a maximum of $n(k+m)$ qubits (see Section~\ref{sec:uni}). The number of qubits in the network is conserved and, as such, there is a one-to-one correspondence between the number of outputs at one step and the number of inputs at the next step. The connections between the outputs and inputs are arbitrary. As such, for a system of $n(k+m)$ qubits there are $(n(k+m))!$ possible connections (referred to as \textit{wirings}) between the outputs of one step and the inputs of the next. Evolution of the network is via the generalized bit oracle defined in Equation~\ref{eq:evolution}.

Consider again a network consisting of an AND and two OR gates. If this network is implemented on a classical system as in Figure~\ref{fig:classcirc}, there are thirty-six possible wiring diagrams and thus thirty-six possible input combinations. However, since each output from the previous step is copied, the number of \textit{unique} input combinations is only eighteen. But in the quantum case as in Figure~\ref{fig:quantcirc}, there are eighty-one possible combinations.

In this article I focus on networks such as these that implement \textit{classical} Boolean functions on a \textit{quantum} network, i.e. a network that allows for quantum inputs and is implemented as described in Section~\ref{sec:circuits}, but I emphasize that the definition introduced here is far more broad.

As an example, I now describe the properties of qABNs of classical functions for the case in which $k=2$. In this case the total number of ancillas $m$ never exceeds the number of functions $n$ since there is never more than one ancilla per function.

\subsection{qABNs of classical functions for $k=2$ inputs}
\label{sec:qabns}
Classical Boolean functions for which $k=2$ and $m=1$ consist of a set of sixteen functions that include the familiar AND, OR, and NOT functions. The most remarkable feature of these networks is the spontaneous order that they exhibit through short state cycles, multiple basins of attraction, and frozen cores separating isolated islands. The most conspicuous differences in $k=2$ \textit{quantum} networks are that the state cycles are not always short (in fact they can be exceptionally long) and, despite the existence of frozen cores and thus percolating walls, the islands are no longer isolated to perturbations.

Simulations of qABNs of these functions in Python utilizing the QuTIP package~\cite{Johansson:2013aa,Johansson:2012aa} have been carried out and a GitHub repository of the code is available (\url{https://github.com/iantdurham/quantum_boolean_networks}). In these simulations the allowed input states were
\begin{equation}
\begin{array}{rlcrl}
\rho(0) & = \ket{0}\bra{0} & & \rho(1) & = \ket{1}\bra{1} \nonumber \\
\rho(+) & = \ket{+}\bra{+} & & \rho(-) & = \ket{-}\bra{-}.
\end{array}
\label{eq:instates}
\end{equation}
Either a set of functions is manually entered or the code randomly chooses a set. Since there are a large number of possible wiring diagrams for any given set of functions, the code is designed to select one by randomly permuting the outputs from the first step. That is, an initial state is set (either randomly chosen or manually entered) and the generalized bit oracle is applied to get an output state which is then randomly permuted before the next application of the oracle. The same permutation is then applied after each subsequent output so that the same wiring occurs at every step. 

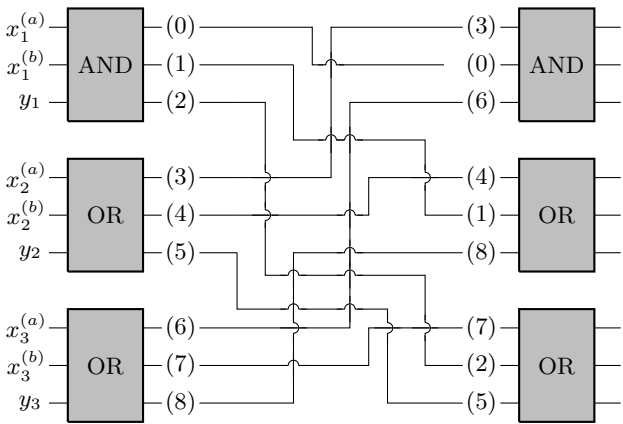
\begin{figure}
\begin{center}
\begin{tikzpicture}
\fill[lightgray] (0,0) -- (1,0) -- (1,1.5) -- (0,1.5) -- (0,0);
\draw[thick] (0,0) -- (1,0) -- (1,1.5) -- (0,1.5) -- (0,0);
\draw (-0.25,0.25) -- (0,0.25);
\draw (-0.25,0.75) -- (0,0.75);
\draw (-0.25,1.25) -- (0,1.25);
\node at (0.5,0.75) {OR};

\node at (-0.55,1.25) {$x_3^{(a)}$};
\node at (-0.55,0.75) {$x_3^{(b)}$};
\node at (-0.5,0.25) {$y_3$};

\draw (1,0.25) -- (1.25,0.25);
\node at (1.5,0.25) {(8)};
\draw (1.75,0.25) -- (2.25,0.25);

\draw (1,0.75) -- (1.25,0.75);
\node at (1.5,0.75) {(7)};
\draw (1.75,0.75) -- (2.25,0.75);

\draw (1,1.25) -- (1.25,1.25);
\node at (1.5,1.25) {(6)};
\draw (1.75,1.25) -- (2.25,1.25);

\fill[lightgray] (0,2) -- (1,2) -- (1,3.5) -- (0,3.5) -- (0,2);
\draw[thick] (0,2) -- (1,2) -- (1,3.5) -- (0,3.5) -- (0,2);
\draw (-0.25,2.25) -- (0,2.25);
\draw (-0.25,2.75) -- (0,2.75);
\draw (-0.25,3.25) -- (0,3.25);
\node at (0.5,2.75) {OR};

\node at (-0.55,3.25) {$x_2^{(a)}$};
\node at (-0.55,2.75) {$x_2^{(b)}$};
\node at (-0.5,2.25) {$y_2$};

\draw (1,2.25) -- (1.25,2.25);
\node at (1.5,2.25) {(5)};
\draw (1.75,2.25) -- (2.25,2.25);

\draw (1,2.75) -- (1.25,2.75);
\node at (1.5,2.75) {(4)};
\draw (1.75,2.75) -- (2.25,2.75);

\draw (1,3.25) -- (1.25,3.25);
\node at (1.5,3.25) {(3)};
\draw (1.75,3.25) -- (2.25,3.25);

\fill[lightgray] (0,4) -- (1,4) -- (1,5.5) -- (0,5.5) -- (0,4);
\draw[thick] (0,4) -- (1,4) -- (1,5.5) -- (0,5.5) -- (0,4);
\draw (-0.25,4.25) -- (0,4.25);
\draw (-0.25,4.75) -- (0,4.75);
\draw (-0.25,5.25) -- (0,5.25);
\node at (0.5,4.75) {AND};

\node at (-0.55,5.25) {$x_1^{(a)}$};
\node at (-0.55,4.75) {$x_1^{(b)}$};
\node at (-0.5,4.25) {$y_1$};

\draw (1,4.25) -- (1.25,4.25);
\node at (1.5,4.25) {(2)};
\draw (1.75,4.25) -- (2.25,4.25);

\draw (1,4.75) -- (1.25,4.75);
\node at (1.5,4.75) {(1)};
\draw (1.75,4.75) -- (2.25,4.75);

\draw (1,5.25) -- (1.25,5.25);
\node at (1.5,5.25) {(0)};
\draw (1.75,5.25) -- (2.25,5.25);

\fill[lightgray] (6,0) -- (7,0) -- (7,1.5) -- (6,1.5) -- (6,0);
\draw[thick] (6,0) -- (7,0) -- (7,1.5) -- (6,1.5) -- (6,0);
\draw (5.75,0.25) -- (6,0.25);
\draw (5.75,0.75) -- (6,0.75);
\draw (5.75,1.25) -- (6,1.25);
\node at (6.5,0.75) {OR};
\draw (7,0.25) -- (7.35,0.25);
\draw (7,0.75) -- (7.35,0.75);
\draw (7,1.25) -- (7.35,1.25);

\node at (5.5,0.25) {(5)};
\node at (5.5,0.75) {(2)};
\node at (5.5,1.25) {(7)};

\fill[lightgray] (6,2) -- (7,2) -- (7,3.5) -- (6,3.5) -- (6,2);
\draw[thick] (6,2) -- (7,2) -- (7,3.5) -- (6,3.5) -- (6,2);
\draw (5.75,2.25) -- (6,2.25);
\draw (5.75,2.75) -- (6,2.75);
\draw (5.75,3.25) -- (6,3.25);
\node at (6.5,2.75) {OR};
\draw (7,2.25) -- (7.35,2.25);
\draw (7,2.75) -- (7.35,2.75);
\draw (7,3.25) -- (7.35,3.25);

\node at (5.5,2.25) {(8)};
\node at (5.5,2.75) {(1)};
\node at (5.5,3.25) {(4)};

\fill[lightgray] (6,4) -- (7,4) -- (7,5.5) -- (6,5.5) -- (6,4);
\draw[thick] (6,4) -- (7,4) -- (7,5.5) -- (6,5.5) -- (6,4);
\draw (5.75,4.25) -- (6,4.25);
\draw (5.75,4.75) -- (6,4.75);
\draw (5.75,5.25) -- (6,5.25);
\node at (6.5,4.75) {AND};
\draw (7,4.25) -- (7.35,4.25);
\draw (7,4.75) -- (7.35,4.75);
\draw (7,5.25) -- (7.35,5.25);

\node at (5.5,4.25) {(6)};
\node at (5.5,4.75) {(0)};
\node at (5.5,5.25) {(3)};

\draw (2.25,3.25) -- (3.5,3.25) -- (3.5,5.25) -- (5.25,5.25);
\node at (3.5,4.75) [jump crossing,scale=1.25]{};

\draw (3.6,4.75) -- (5,4.75);
\draw (2.25,5.25) -- (3.25,5.25) -- (3.25,4.75) -- (3.4,4.75);

\draw (2.25,1.25) -- (3.75,1.25) -- (3.75,4.25) -- (5.25,4.25);
\draw (2.25,2.75) -- (3.65,2.75);
\node at (3.75,2.75) [jump crossing,scale=1.25]{};
\draw (3.85,2.75) -- (4,2.75) -- (4,3.25) -- (5.25,3.25);

\draw (2.25,4.75) -- (3,4.75);
\node at (3.75,3.75) [jump crossing,scale=1.25]{};
\node at (3.5,3.75) [jump crossing,scale=1.25]{};
\draw (3,4.75) -- (3,3.75) -- (3.4,3.75);
\draw (3.85,3.75) -- (4.75,3.75) -- (4.75,3.35);
\node at (4.75,3.25) [jump crossing,rotate=-90,scale=1.25]{};
\draw (4.75,3.15) -- (4.75,2.75) -- (5.25,2.75);

\draw (2.25,0.25) -- (3,0.25) -- (3,1.15);
\node at (3,1.25) [jump crossing,rotate=-90,scale=1.25]{};
\draw (3,1.35) -- (3,2.25) -- (3.65,2.25);
\node at (3.75,2.25) [jump crossing,scale=1.25]{};
\draw (3.85,2.25) -- (5.25,2.25);

\draw (2.25,0.75) -- (2.9,0.75);
\node at (3,0.75) [jump crossing,scale=1.25]{};
\draw (3.1,0.75) -- (4,0.75) -- (4,1.25) -- (5.25,1.25);

\draw (2.25,2.25) -- (2.25,1.5) -- (2.9,1.5);
\node at (3.75,1.5) [jump crossing,scale=1.25]{};
\draw (3.1,1.5) -- (3.65,1.5);
\node at (3,1.5) [jump crossing,scale=1.25]{};
\draw (3.85,1.5) -- (4.25,1.5) -- (4.25,1.35);
\node at (4.25,1.25) [jump crossing,rotate=-90,scale=1.25]{};
\draw (4.25,1.15) -- (4.25,0.25) -- (5.25,0.25);

\draw (2.25,4.25) -- (2.625,4.25) -- (2.625,3.35);
\node at (2.625,3.25) [jump crossing,rotate=-90,scale=1.25]{};
\draw (2.625,3.15) -- (2.625,2.85);
\node at (2.625,2.75) [jump crossing,rotate=-90,scale=1.25]{};
\draw (2.625,2.65) -- (2.625,1.95) -- (2.9,1.95);
\node at (3.75,1.95) [jump crossing,scale=1.25]{};
\draw (3.1,1.95) -- (3.65,1.95);
\node at (3,1.95) [jump crossing,scale=1.25]{};
\draw (3.85,1.95) -- (4.75,1.95) -- (4.75,1.35);
\node at (4.75,1.25) [jump crossing,rotate=-90,scale=1.25]{};
\draw (4.75,1.15) -- (4.75,0.75) -- (5.25,0.75);

\end{tikzpicture}
\end{center}
\caption{\label{fig:wiring} The wires exiting the first step, $[0,1,2,3,4,5,6,7,8]$, enter the second step in the order $[3,0,6,4,1,8,7,2,5]$, i.e. the circuit maps $x_{1,t}^{(a)}\mapsto x_{1,t+1}^{(b)}$.}
\end{figure}

In order to keep track of a specific wiring diagram, I assign an index to the output `wires' for a network starting with 0 at the top. Each wire can then be traced to its input at the next step which shuffles the indices. For the network shown in Figure~\ref{fig:quantcirc}, there are nine output wires that are labeled $[0,1,2,3,4,5,6,7,8]$ top-to-bottom in the figure. The order in which these wires connect to the next iteration of the network is $[3,0,6,4,1,8,7,2,5]$ as shown in Figure~\ref{fig:wiring} (e.g. the circuit maps $x_{1,t}^{(a)}\mapsto x_{1,t+1}^{(b)}$, etc.).

Tracking the state cycles for quantum networks can be particularly tricky given the shear number of possible states that could exist in a given cycle. The addition of ancillas adds to the complexity level. We can, however, get some sense of the behavior of the system as a whole by tracking various properties of the system. For example, one measure that provides a sense of how these networks evolve is the multipartite mutual information which, for a network whose overall state is $\rho$, is given as~\cite{Watanabe:1960aa,Herbut:2004aa}
\begin{align}
I_m(A_1 : \cdots : A_n) & = \sum_n S(\rho^{A_n}) - S(\rho) \\ \nonumber
& = S(\rho || \rho^{A_1} \otimes \cdots \otimes \rho^{A_n}) \\ \nonumber
& \ge 0
\label{qmi}
\end{align}
where $S(\rho)$ is the von Neumann entropy of the entire network and $S(\rho^{A_n})$ is the von Neumann entropy of the $n$th qubit in the network. It expresses the difference between the sum of the von Neumann entropies of the individual qubits in the system, and the von Neumann entropy of the system considered as a whole. Since the network evolves deterministically in the absence of measurement, the value $I_m$ should vary periodically. Note, however, that this measure does not depend on the relative locations of the qubits in the network. As such it cannot distinguish between, say, the state $\rho(+) \otimes \rho(-) \otimes \rho(0)$ and the state $\rho(-) \otimes \rho(+) \otimes \rho(0)$. But there exists a $k=2$ Boolean function for which $f(\rho(+) \otimes \rho(-) \otimes \rho(0))\ne f(\rho(-) \otimes \rho(+) \otimes \rho(0))$ and thus the order of the qubits matters. As such, the length of any cycles for $I_m$ would represent a \textit{minimum} length for any associated \textit{state} cycles. That is, if, for example, $I_m$ for a given network varies over five time steps, the state cycle for the network could be no shorter than five but might be longer. Developing a more robust measure is an important aim of future work. However, a fair amount of useful information can be obtained about a network from the behavior of multipartite mutual information as I will show.

\begin{figure}
\includegraphics[width=0.5\textwidth]{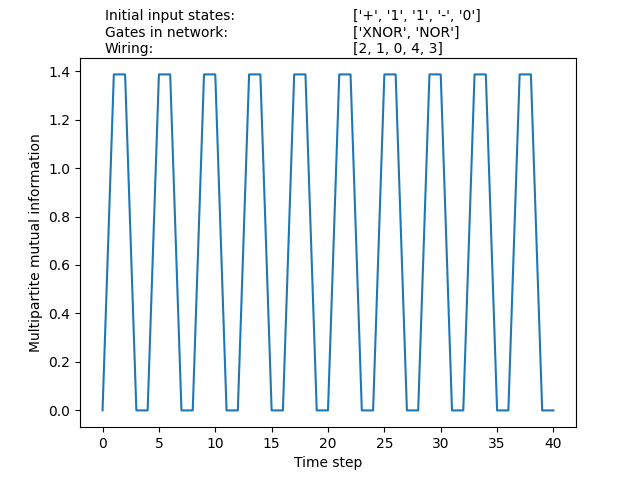}
\caption{\label{fig:sym} A quantum network implementing an XNOR function and a NOR function with an input state of $\rho(\mathbf{X},\mathbf{Y})=\rho(\mathbf{x}_1,\mathbf{y}_1)\otimes \rho(\mathbf{x}_2,\mathbf{y}_2)=\rho(+,1)\otimes\rho(1-,0)$ and wiring $[2,1,0,4,3]$ has a multipartite mutual information that displays symmetric oscillatory behavior with a cycle length of six steps.}
\end{figure}

\begin{figure}
\includegraphics[width=0.5\textwidth]{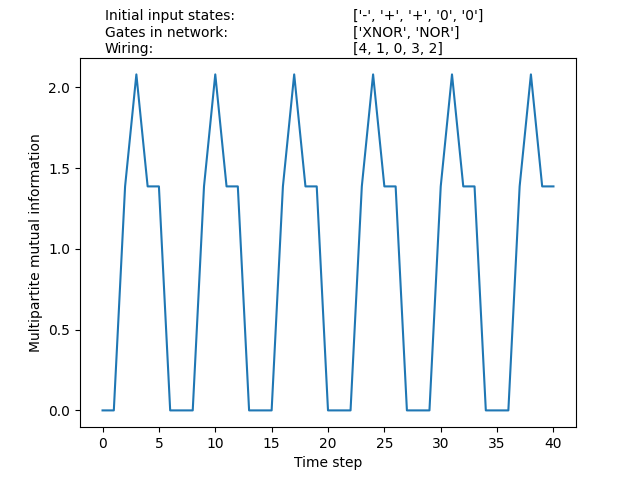}
\caption{\label{fig:nonsym} A network implementing an XNOR and a NOR function with an input state of$\rho(\mathbf{X},\mathbf{Y})=\rho(\mathbf{x}_1,\mathbf{y}_1)\otimes \rho(\mathbf{x}_2,\mathbf{y}_2)=\rho(-,+)\otimes\rho(+0,0)$ and wiring $[4,1,0,3,2]$ exhibits asymmetric behavior.}
\end{figure}

As a simple example of the use of the multipartite mutual information for analyzing these networks, consider a network consisting XNOR and NOR functions with a wiring of $[0,3,2,1,4]$ and an initial input state of $\rho(\mathbf{X},\mathbf{Y})=\rho(\mathbf{x}_1,\mathbf{y}_1)\otimes \rho(\mathbf{x}_2,\mathbf{y}_2)=\rho(+,+)\otimes\rho(10,0)$. In this case the multipartite mutual information, $I_m$, varies symmetrically with a cycle length of six steps as shown in Figure~\ref{fig:sym}.
While the actual state cycle length may be longer, the cycle length for $I_m$ nevertheless is useful in that it suggests that the network evolves through the states in the cycle in a symmetric way. In contrast, a network implementing the same functions but with input state $\rho(\mathbf{X},\mathbf{Y})=\rho(\mathbf{x}_1,\mathbf{y}_1)\otimes \rho(\mathbf{x}_2,\mathbf{y}_2)=\rho(-,+)\otimes\rho(+0,0)$ and wiring $[4,1,0,3,2]$ exhibits asymmetric behavior, as shown in Figure~\ref{fig:nonsym}.

\begin{figure}
\includegraphics[width=0.5\textwidth]{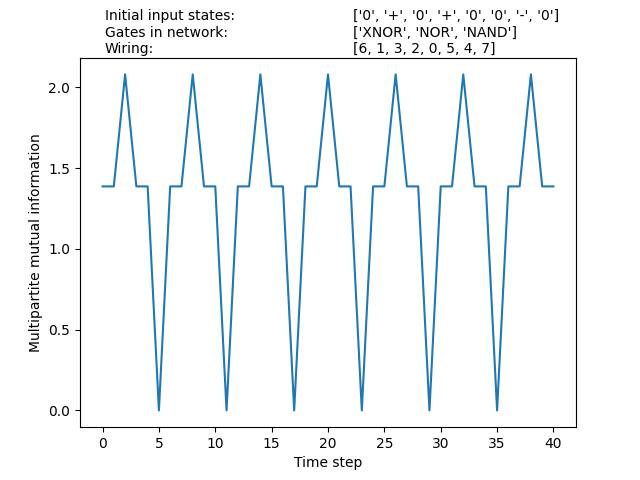}
\caption{\label{fig:XNN_short} A network implementing XNOR, NOR, and NAND functions with an input state of $\rho(\mathbf{X},\mathbf{Y})=\rho(\mathbf{x}_1,\mathbf{y}_1)\otimes\rho(\mathbf{x}_2,\mathbf{y}_2)\otimes\rho(\mathbf{x}_3,\mathbf{y}_3)=\rho(0,+)\otimes\rho(-+,0)\otimes\rho(0-,0)$ and wiring $[6,1,3,2,0,5,4,7]$ exhibits a relatively short cycle length for $I_m$.}
\end{figure}

\begin{figure}
\includegraphics[width=0.5\textwidth]{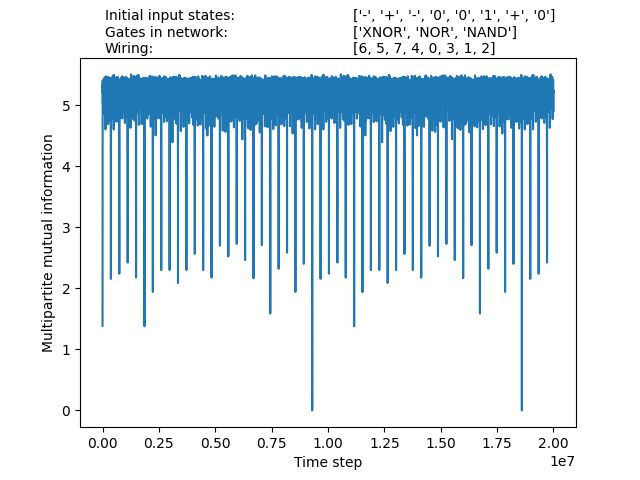}
\caption{\label{fig:XNN_long} A network implementing XNOR, NOR, and NAND functions with an input state of $\rho(\mathbf{X},\mathbf{Y})=\rho(\mathbf{x}_1,\mathbf{y}_1)\otimes\rho(\mathbf{x}_2,\mathbf{y}_2)\otimes\rho(\mathbf{x}_3,\mathbf{y}_3)=\rho(-+,0)\otimes\rho(-+,0)\otimes\rho(-,1)$ and wiring $[6,3,1,4,5,7,2,0]$ exhibits a very long cycle length for $I_m$.}
\end{figure}

One might expect the complexity of the behavior exhibited by these networks to increase rapidly as the number of functions increases. But suppose we add a single NAND function to the simple network we just considered. That is, suppose we implement a network implementing XNOR, NOR, and NAND functions with an input state of $\rho(\mathbf{X},\mathbf{Y})=\rho(\mathbf{x}_1,\mathbf{y}_1)\otimes\rho(\mathbf{x}_2,\mathbf{y}_2)\otimes\rho(\mathbf{x}_3,\mathbf{y}_3)=\rho(0,+)\otimes\rho(-+,0)\otimes\rho(0-.0)$ and wiring $[6,1,3,2,0,5,4,7]$. The cycle length for $I_m$, shown in Figure~\ref{fig:XNN_short}, is just one step longer than the associated cycle for the network in Figure~\ref{fig:sym}.
By contrast, the same set of functions with an input state of $\rho(\mathbf{X},\mathbf{Y})=\rho(\mathbf{x}_1,\mathbf{y}_1)\otimes\rho(\mathbf{x}_2,\mathbf{y}_2)\otimes\rho(\mathbf{x}_3,\mathbf{y}_3)=\rho(-,+)\otimes\rho(-0,0)\otimes\rho(1+,0)$ and wiring $[6,5,7,4,0,3,1,2]$ has a cycle length for $I_m$ of nearly nine million steps, as shown in Figure~\ref{fig:XNN_long}.
This actually tells us something about the states in the cycle. We know that the state cycle must be as long if not longer than the cycle length of $I_m$ and, given that there are eight qubits in the system, if those qubits could only ever be in one of the four input states given by equation~\ref{eq:instates}, there would only be a total of $4^8$ possible system states. Since the system is deterministic in that each state uniquely leads to another state, $4^8$ would thus be the maximum length for any state cycle. Since the state cycle for the network in Figure~\ref{fig:XNN_long} far exceeds that, the network \textit{must produce states not specified by equation}~\ref{eq:instates}. As such, while $I_m$ cannot distinguish between all of the network's states, it can still tell us something important about what the network is doing and can prompt further action. In this example, when tracing out the individual qubit states, one finds that they are mixed states.

The network described in Figure~\ref{fig:XNN_short} also exhibits two frozen cores. Since the $y_1$ (index 1), $x_3^{(a)}$ (index 5), and $y_3$ (index 7) qubits feed back into themselves, one might actually expect there to be three frozen cores. But, in fact, only $y_1$ and $x_3^{(a)}$ remain unchanged. It might be unsurprising that $x_3^{(a)}$ is frozen given that no output is mapped to it. But $y_1$ represents the logical output of the XNOR function, i.e. $y_1^{\prime} = y_1 \oplus f_1$. Thus, if $x_1$ changes, one would expect $y_1$ to change. If we look at the states that $x_1$ cycles through --- $\rho(-)$, $\rho(0)$, and a mixed state defined as $0.5\rho(0)+0.5\rho(1)$ --- we might expect that, at the very least, $y_1$ would change to a mixed state of some kind, but the combinations are just such that no change occurs.

The frozen cores $y_1$ and $x_3^{(a)}$ separate the network into three ``islands'' of states: one island on the left containing $x_1$, one island in the middle containing $x_2^{(a)}$, $x_2^{(b)}$, and $y_2$, and one island on the right containing $x_3^{(b)}$ and $y_3$. In classical Boolean networks, perturbations to bits in one island have no effect on the bits in the other islands. Hence the islands are said to be \textit{isolated}. The same is not true in the quantum case. As an example, consider that on the eleventh time step, $x_1=\rho(0)$ and on the fifteenth time step, $y_2=\rho(0)$. If we apply the bit flip operator to $x_1$ on the eleventh time step, $y_2$ at the fifteenth time step also flips. In a classical network, $y_2$ would never flip since it is in a different island. One would assume that the ability to cross into another island is a result of non-classical correlations, but a better measure is needed in order to be sure.

But this result is remarkable for another reason: $y_2$ is not affected until the fifteenth step. Perhaps even more remarkably, this perturbation has no effect whatsoever on the multipartite mutual information cycle. While this can be (rightly) taken as a failure of $I_m$ to detect changes to the individual qubit states, it is nevertheless notable that $I_m$ is resilient to perturbations in this situation. In the language of classical Boolean networks, the cycle would be referred to as an \textit{attractor} since the perturbation ultimately preserved the cycle. Not all cycles (classical or quantum) are impervious to perturbations.

Having set out the basic definition of qABNs and given a few specific examples, it's now worth taking a closer look at some of the pressing questions and how they might be addressed.

\section{Discussion and Conclusions}
\label{sec:conc}
Perhaps the most pressing issue raised by this framework is the need for an alternative measure to the multipartite mutual information that can distinguish states in such a way as to maintain the ordering of the individual qubits since it is what distinguishes the the states in the cycle. One potential measure for this is the quantum discord~\cite{Ollivier:2001,Henderson:2001aa,Zurek:2002} since the discord can be asymmetric~\cite{Dill:2008,Dak:2010}. However, existing generalizations of the discord to multipartite systems are symmetric with regard to the exchange of subsystems~\cite{Rulli:2011,Rad:2020}, though the latter can account for different measurement orders. If the latter could be further extended to account for the ordering of the qubits, it might be useful, particularly since it tells us something about the correlations between the various subsystems. However, it likely would also prove to be difficult to compute since computing the discord is known to be NP-complete~\cite{Huang:2011aa}.

Another pressing issue is the nature of the mixed states that are produced in some of these networks. In the network shown in Figure~\ref{fig:XNN_short}, recall that a perturbation to $x_1$ on the eleventh time step led to a change in the value of $y_2$ on the fifteenth time step. One would expect that a perturbation at one time step would lead to changes at the very next time step rather than several steps later. It seems possible that the propagation of the perturbation occurs through some combination of correlations between the states as well as the nature of the mixed states. That is, the mixed states somehow mask the perturbation. A better understanding of these mixed states, particularly in conjunction with a better measure than the multipartite mutual information, might shed more light on this issue.

\begin{figure}
\includegraphics[width=0.5\textwidth]{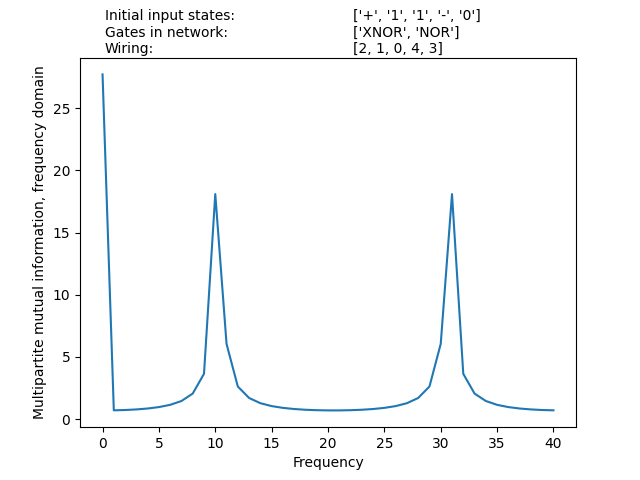}
\caption{\label{fig:transform} The Fourier transform of $I_m$ for the network shown in Figure~\ref{fig:sym} likely represents the Fourier decomposition of the cycle, but it's unclear which subsystems the components would represent.}
\end{figure}

Given that these networks exhibit oscillatory behavior, as evidenced by their finite state cycles, it might also be of interest to explore any cycle measure such as the multipartite mutual information or discord in the frequency domain. In fact the code written for this and available on GitHub includes a Fourier transform of the multipartite mutual information to the frequency domain. The transform for the cycle shown in Figure~\ref{fig:sym} is given in Figure~\ref{fig:transform}.
The trouble is that it's not entirely clear how to interpret the results. One would expect that it would give the Fourier decomposition of a given cycle into component cycles, but it's not clear what those component cycles would represent. Are they, for instance, the cycles of individual subsystems within the larger network? If so, which subsystems would they represent?

In addition to these particular questions, I have only explored networks of $k=2$ functions. Additional networks should be explored in order to establish general bounds for the mean attractor length and mean attractor number for each class of network.

Nevertheless, the framework presented here captures some of the most salient features of classical Boolean networks in a quantum setting including finite state cycles and frozen cores, while also demonstrating uniquely quantum properties within a rich landscape of behavior ripe for further exploration.

\acknowledgements
I would like to thank Nana Liu, Peter Rohde, Robert Prentner, and Larissa Albantakis for helpful discussions. This work was partially supported by a grant from FQxI (FQXi-RFP-IPW-1911).

\bibliographystyle{apsrev}
\bibliography{qboolean.bib}
\end{document}